\documentclass[square,comma,sort&compress]{elsart}
\usepackage{amsfonts}
\usepackage{amsmath}
\usepackage{natbib}
\usepackage{epsfig}
\usepackage{rotating}
\usepackage{graphicx}

\newcommand{\QUESTION}[1]{}
\newcommand{\ANSWER}[1]{}

\begin{document}

\runauthor{Tessarotto}
\begin{frontmatter}
\title{Microscopic statistical description of classical matter around black holes}
\author[UTS,TO,TS2,LJ]{P.Nicolini\thanksref{Someone}}
\author[UTS,TS3]{and M. Tessarotto}

\address[UTS]{Department of Mathematics and Informatics,
University\,of\,Trieste, Trieste, Italy}
\address[TO]{Department of Mathematics,
Polytechnical University\,of\, Turin,\ Turin, Italy}
\address[TS2]{INFN, National Institute for Nuclear Physics, Trieste section, Trieste, Italy}
\address[TS3]{Consortium for Magnetofluid Dynamics, Trieste,Italy}
\address[LJ]{Jo\v{z}ef Stefan Institute, Ljubljana, Slovenia}
\thanks[Someone]{Corresponding author: email:
Piero. Nicolini@cmfd.univ.trieste.it}
\begin{abstract}
The problem of statistical description of classical matter,
represented by a N-body system of  relativistic point particles
falling into a black hole, is investigated, adopting a classical
framework.  A  covariant microscopic statistical description is
developed for the classical particle system by introducing the
definition of observer-related N-particle microscopic distribution
function and by suitably extending to general relativity the
concept of Gibbs microscopic entropy. The corresponding entropy
production rate is calculated taking into account the presence of
an event horizon  and proven to obey, for $N>>1$, an exact
microscopic H-theorem which holds for arbitrary geometries.
\end{abstract}

\end{frontmatter}

\section{Introduction}

Since its inception \cite{Bekenstein 1973,Bekenstein 1974,Hawking
1975,Bekenstein 1975,Hawking 1976} the thermodynamical
interpretation of black holes (BH) has been the subject of debate
\cite{Fursaev 2004}. Indeed, the mathematical analogy between the
laws of thermodynamics and black hole physics following from
classical general relativity still escapes a complete and
satisfactory interpretation. In particular it is not yet clear
whether this analogy is merely formal or leads to an actual
identification of physical quantities belonging to apparently
unrelated frameworks. The difficulty is related to the very
concept of entropy usually adopted in BH theory, based on
Boltzmann entropy, which is determined by the number $W$ of
microscopic complexions compatible with the macroscopic state of a
physical system
\begin{equation*}
S_{bh}=K\ln W.
\end{equation*}%
Indeed $S_{bh}$ does not rely on a true statistical description of
physical systems, but only on the classification of the
microstates of the system, quantal or classical. As a consequence,
the evaluation of $S_{bh}$ requires the knowledge of the internal
structure of the BH, a result which obviously cannot be achieved
in the context of a purely classical description of BH. Therefore
the evaluation of $S_{bh}$ requires a consistent formulation of
quantum theory of gravitation and matter \cite{Bekenstein
1975,Hawking 1976}. This can be based, for example, on string
theory \cite{Reviews 1998} and can be conveniently tested in the
framework of semiclassical gravity \cite{Nicolini 2001,Nicolini
2002}.

A basic difficulty of quantum theories founded on the concept of
Boltzmann entropy, is that up not now they have not leaved up to
their expectations since they have not yet achieved their primary
goal, i.e., the rigorous proof of an H-theorem and the full
consistency with the second law of thermodynamics $\delta
S_{bh}\geq 0$. Indeed, estimates of the Boltzmann entropy based
quantum theory of gravitation \cite{Bekenstein 1975,Hawking 1976}
and yielding $S_{bh}\equiv \frac{1}{4}k\frac{c^{3}A}{G\hbar },$
being $A$ the area of the event horizon, are inconsistent from
this viewpoint, since as a consequence of the BH radiation effect
\cite{Hawking 1975} the radius of the BH may actually decrease.
Hence, the Boltzmann entropy $S_{bh}$ \textit{cannot be
interpreted}, in a proper sense, as a physical entropy of the BH.
To resolve this difficulty a modified constitutive equation for
the entropy was postulated \cite{Bekenstein 1973,Bekenstein 1974},
in order to include the contribution of the matter in the BH
exterior, by setting
\begin{equation}
S^{\prime }=S+S_{bh},  \label{Beck}
\end{equation}%
($S^{\prime }$ denoting the so-called \textit{Bekenstein entropy}) where $S$
represents the correction carried by the matter outside the BH (notice,
however, that also $S$ cannot be interpreted as entropy). As a consequence a
generalized second law $\delta S^{\prime }\geq 0$ was proposed \cite%
{Bekenstein 1973,Bekenstein 1974} which can be viewed as nothing more than
the ordinary second law of thermodynamics applied to a system containing a
BH. However, the precise definition and underlying statistical basis both
for $S$ and potentially also of $S_{bh}$ remain obscure. Thus a fundamental
problem appears their precise estimates based on suitable microscopic models.

On the other hand, if one regards the BH as a classical object in
the space-time continuum, provided the surrounding falling matter
can be assumed as formed by a suitably large number of particles,
the estimate of the BH entropy should be achievable directly in
the context of classical statistical mechanics, by adopting the
customary concept of statistical entropy, i.e., Gibbs entropy. In
contrast to the Boltzmann entropy, this is based on a statistical
description of physical systems and is defined in terms of the
probability distribution of the observable microstates of the
system. In fact, as is well-known, its definition coincides with
the axiomatic definition of Shannon entropy, yielding the measure
of ignorance on statistical ensembles. \ A first result of this
type been presented in previous work by Nicolini and Tessarotto
\cite{Nicolini2005a} (hereon denoted as Ref.I), where a covariant
kinetic theory in the presence of an event horizon was developed
for classical matter in the BH exterior, treated as a N-body
system (with $N\gg 1$) of classical point particles undergoing,
before capture, a purely Hamiltonian dynamics. By introducing a
suitable definition for the relativistic kinetic entropy of
infalling classical matter described as an ensemble of point
particles, an H-Theorem was reached, tanking into account the
presence of the black hole event horizon.

The goal of the present paper is to extend the results of Ref.I by
developing a microscopic (N-body) statistical description of
classical matter in the presence of an event horizon.
For this purpose we intend to evaluate the Gibbs
(microscopic) entropy of classical matter falling into the BH
event horizon. \ In particular, we intend to prove the validity of
an exact H-theorem holding for Gibbs entropy, provided the
infalling matter can be described as a suitably large classical
system ($N\gg 1$) of point (neutral or charged) particles forming
plasma or a rarefied gas which interact mutually only via a mean
Hamiltonian field. \

The scheme of the presentation is as follows. In Sec. II, first the basic
assumptions of the theory are introduced, which include the concept of
particle capture domain, assumed to be defined by a smooth hypersurface
located suitably close to the event horizon (\textit{capture surface}).
Second, a covariant microscopic statistical description is formulated for a
N-body system of classical point particles referred to an arbitrary
observer. In particular, in the exterior domain of the BH (outside the
capture surface) the particle system is assumed to obey a Hamiltonian
dynamics. As a consequence, in the same domain the Liouville theorem for the
corresponding phase-flow is readily recovered and, based on the concept of
observer-related N-particle microscopic distribution function and
probability density, the relativistic N-body Liouville equation is
determined. Finally suitable boundary conditions are introduced for the
microscopic distribution function on the capture surface.

In Sec. III the definitions of observer-related relativistic Gibbs and
kinetic entropies for the infalling classical matter are introduced and the
corresponding entropy production rates are determined. The relationship
between the two entropies is proven to hold in an elementary way for
classical systems of weakly interacting point particles. As a consequence
the obvious physical interpretation of the contributions of the entropy
production rate are pointed out.

Finally, in Sec. IV, an H-theorem is recovered for the Gibbs entropy,
extending the result of Ref.I. The result is proven to hold for BH
characterized by an event horizon of arbitrary shape and size. In
particular, the result is proven to apply, in contrast to Boltzmann entropy
in quantum theory of gravitation \cite{Hawking 1976}, also to the case of
matter falling in a BH with contracting event horizons, such as BH
implosions or slow contractions.

\section{The N-body covariant microscopic statistical description}

In this section we introduce the basic framework of the theory.
The assumptions we are going to make deal with the treatment of
non-isolated black hole formed by the collapse of a star and
surrounded by matter. The BH, together with its associated event
horizon, and of the matter surrounding the BH and falling toward
the event horizon are all assumed as classical. It is sufficient
for our purposes to assume the following hypotheses:

\textit{I) Infalling particles capture is due to the redshift
phenomenon occurring near the event horizon for an arbitrary
observer located far from the BH (for
example in a region where space time is asymptotically flat \cite{Wald 1984}%
). As a result particles sufficiently close to the event horizon
effectively disappear to the observer (Assumption 1).} In the
sequel we shall assume that all particle capture events occurs in
a subdomain, to be identified with a surface $\gamma $ of the
space-time (\textit{capture surface}), localized infinitesimally
close to the event horizon.

\textit{II) The total energy of infalling particles is finite in
such a way that local distortions of space-time are negligible
(Assumption 2). }

The matter outside the BH is described by a system of $N\gg 1$ identical
classical particles to be referred to an arbitrary observer $O$. If we
assume that the system is Hamiltonian, its evolution is well known and
results uniquely determined by the classical equations of motion, defined
with respect to the observer $O$. To this purpose let us choose $O$, without
loss of generality, in a region where space time is (asymptotically) flat,
endowing it with the proper time $\tau $, where $\tau $ is assumed to span
the set $I\subseteq R$ (observer's time axis). Without loss of generality we
can assume that the particles are points, i.e., they are described by the
1-particle canonical states $\mathbf{x}_{i}$ (with $i=1,N$) spanning the $8-$%
dimensional phase space $\Gamma _{i},$ where ${\mathbf{x}}_{i}=\left(
r_{i}^{\mu },p_{i\mu }\right) $. The analogous treatment of particles having
higher degree of freedom is straightforward. Therefore, the evolution of the
system, described in terms of the N-body canonical state ${\mathbf{x}}\equiv
\left\{ {\mathbf{x}}_{1},...,{\mathbf{x}}_{N}\right\} ,$ is determined by a
suitable relativistic Hamiltonian $H=H\left( \mathbf{x}\right) $, where each
canonical 1-particle state ${\mathbf{x}}_{i}$ $(i=1,N)$ results
parameterized in terms of the $i-$th particle world line arc length $s_{i}$
(see \cite{Synge 1960}). As a consequence, requiring that $s_{i}=s_{i}(\tau
) $ results a strictly monotonic function it follows that, the particle
state can be also parameterized in terms of the observer's time $\tau .$
Therefore, the particle states are determined by the canonical equations:%
\begin{eqnarray}
\frac{ds_{(i)}}{d\tau }\frac{d\mathbf{x}_{i}(s_{i}(\tau ))}{ds_{(i)}} &=&%
\frac{ds_{(i)}}{d\tau }\left[ \mathbf{x}_{(i)},H\right] _{\mathbf{x}_{i}},
\label{Hamiltonian dynamics} \\
\mathbf{x}_{i}(s_{i}(\tau _{o})) &=&\mathbf{x}_{oi},  \notag
\end{eqnarray}%
where $\left[ f(\mathbf{x}_{(i)}),H\right] _{\mathbf{x}_{i}}$ is the
canonical Poisson equation%
\begin{equation}
\left[ f(\mathbf{x}_{(i)}),H\right] _{\mathbf{x}_{i}}=\left\{ \frac{\partial
f(\mathbf{x}_{(i)})}{\partial r_{i}^{\mu }}\frac{dH}{dp_{(i)\mu }}-\frac{%
\partial f(\mathbf{x}_{(i)})}{\partial p_{i\mu }}\frac{dH}{dr_{(i)}^{\mu }}%
\right\}
\end{equation}%
and the previous initial-value problem admits by assumption a unique
solution defining a $C^{(2)}(\Gamma ^{N}\times I)$-solution. It follow that
the phase-flow defined by the mapping $\mathbf{x}_{o}=\left\{ \mathbf{x}%
_{o1},...,\mathbf{x}_{oN}\right\} \rightarrow \mathbf{x}(\tau )\equiv
\left\{ \mathbf{x}_{1}(s_{1}(\tau )),.....,\mathbf{x}_{N}(s_{N}(\tau
))\right\} $ satisfies a Liouville theorem, namely $\frac{d}{d\tau }%
\left\vert \frac{\partial \mathbf{x}(\tau )}{\partial \mathbf{x}_{0}}%
\right\vert =0$ (see Appendix) and therefore the canonical measure
defining the phase space volume
$d\mathbf{x=}\prod\limits_{i=1,N}d\mathbf{x}_{i}$ is conserved,
since $d\mathbf{x}=d\mathbf{x}_{o}$.

Next, to obtain a microscopic statistical description for the N-body system
we introduce the $N-$\textit{body microscopic distribution function for the
observer} $O$, $\rho _{G}^{(N)}(\mathbf{x})$,
\begin{equation}
\rho _{G}^{(N)}({\mathbf{x}})\equiv \rho ^{(N)}({\mathbf{x}}%
)\prod\limits_{i=1,N}\delta (s_{i}-s_{i}(\tau
))\prod\limits_{j=1,N}\delta (\sqrt{u_{j\mu }u_{j}^{\mu
}}-1)=Np_{G}^{(N)}({\mathbf{x}})
\end{equation}%
defined on the extended $N-$ particle phase-space $\Gamma
^{N}=\prod\limits_{i=1,N}$ $\Gamma ^{1,i}$ (where $\Gamma ^{1,i}$, for $%
i=1,N$, is the 8-dimensional phase space of the $i$-th particle).$.$ Here $%
\rho ^{(N)}\left( \mathbf{x}\right) $ denotes the conventional microscopic
distribution function in the $8N-$dimensional phase space, while $%
p_{G}^{(N)}({\mathbf{x}})$ is the corresponding $N$-body microscopic
probability density satisfying the normalization%
\begin{equation}
\int\limits_{\Gamma ^{N}}d\mathbf{x}p_{G}^{(N)}({\mathbf{x}})=1.
\end{equation}%
Notice that the Dirac deltas introduced above must be intended as \emph{%
physical realizability equations}. In particular the condition placed on the
arc lengths $s_{i}$ implies that the $i$-th particle of the system is
parameterized with respect to $s_{i}({\tau })$, i.e., it results
functionally dependent on the proper time of the observer; instead the
constraints placed on $4$-velocity implies that $u_{j\mu }$ must belong to
the hypersurface $\delta _{i}$ of equation $\sqrt{u_{i\mu }u_{i}^{\mu }}=1,$
and hence $u_{i}^{\mu }$ is a tangent vector to a timelike geodesic. In the
sequel we adopt also the notation
\begin{equation}
\widehat{\rho }^{(N)}({\mathbf{x}})=\left. \rho ^{(N)}({\mathbf{x}}%
)\right\vert _{\left\{ \sqrt{u_{1\mu }u_{1}^{\mu }}=1,...\sqrt{u_{N\mu
}u_{N}^{\mu }}=1\right\} },  \label{ro cappello}
\end{equation}%
to denote $\rho ^{(N)}({\mathbf{x}})$ evaluated on the intersection of the
hypersurfaces $\delta _{1},....,\delta _{N}.$ The \emph{event horizon} of a
classical BH is defined by the hypersurface $r_{H},$identical for all
particles and specified by the equation
\begin{equation}
R_{m}=r_{H},
\end{equation}%
where for $m=1,N,$ $R_{m}$ denotes the a suitable curvilinear coordinate for
the $m$-th particle and coincides with the radial coordinate in the
spherically symmetric case. According to a classical point of view, let us
now assume that the particle capture surface be defined by the surface $%
\gamma $ of equation
\begin{equation}
R_{m}(s_{m})=r_{\epsilon }(s_{m}),
\end{equation}%
where $r_{\epsilon }=(1+\epsilon )r_{H}$, while $\epsilon >0$ is
an infinitesimal which may depend on explicit or hidden parameters
(for example, $\epsilon $ might depend on the detector used by the
observer). The presence of the BH event horizon is taken into
account by defining suitable boundary conditions for the kinetic
distribution function on the hypersurface $\gamma $. For this
purpose we distinguish between incoming and
outgoing distributions on $\gamma $ with respect to the $i$-th particle, $%
\left. \rho _{G}^{(N)(+,i)}({\mathbf{x})}\right\vert _{\gamma }$ and $\left.
\rho _{G}^{(N)(-,i)}({\mathbf{x}})\right\vert _{\gamma }$ corresponding
respectively to $n_{\alpha }u_{i}^{\alpha }>0$ and $n_{\alpha }u_{i}^{\alpha
}\leq 0$, where $n_{\alpha }$ is a locally radial outward $4-$vector.
Therefore the boundary conditions on $\gamma $ are specified as follows
\begin{eqnarray}
\left. \rho _{G}^{(N)(-,i)}({\mathbf{x})}\right\vert _{\gamma }
&\equiv &\rho ^{(N)}({\mathbf{x}})\prod\limits_{i=1,N}\delta
(s_{i}-s_{i}(\tau
))\prod\limits_{j=1,N}\delta (\sqrt{u_{j\mu }u_{j}^{\mu }}-1), \\
\left. \rho _{G}^{(N)(+,i)}({\mathbf{x}})\right\vert _{\gamma }
&\equiv &0. \notag
\end{eqnarray}%
As previously anticipated, these boundary conditions do not actually require
the detailed physical model for the particle loss mechanism, since all
particles are assumed to be captured on the same hyper-surface $\gamma $,
independent of their state. This provides a classical loss model for the BH
event horizon.

\section{Covariant Liouville equation}

It is now immediate to determine the covariant Liouville equation
which in the external domains, i.e. outside the event horizon,
advances in time the microscopic distribution function \ $\rho
_{G}^{(N)}(\mathbf{x})$ with respect to the observer $O.$ Thanks
to Liouville theorem (i.e., phase-space volume conservation in
$\Gamma ^{N}$ in the sense indicated above) and invoking as usual
the axiom of conservation of probability for classical systems, it
follows that $\widehat{\rho }^{(N)}(\mathbf{x})$ must satisfies
the differential Liouville equation
\begin{equation}
\frac{ds_{i}}{d\tau }\left\{ \frac{dr_{i}^{\mu }}{ds_{i}}\frac{\partial
\widehat{\rho }^{(N)}(\mathbf{x})}{\partial r_{i}^{\mu }}+\frac{dp_{i\mu }}{%
ds_{i}}\frac{\partial \widehat{\rho }^{(N)}(\mathbf{x})}{\partial p_{i\mu }}%
\right\} =0,  \label{Liouville equation}
\end{equation}%
where by assumption
\begin{equation}
\frac{ds_{i}(\tau )}{d\tau }>0  \label{assumption}
\end{equation}%
is made for all $i=1,N,$ and the summation is understood over repeated
indexes. This equation resumes the conservation of the probability in the
relativistic phase space in the domain external to the event horizon.
Invoking the Hamiltonian dynamics (\ref{Hamiltonian dynamics}), the kinetic
equation takes the conservative form
\begin{equation}
\frac{ds_{i}}{d\tau }\left[ \widehat{\rho }^{(N)}(\mathbf{x}),H\right] _{%
\mathbf{x}_{i}}=0.
\end{equation}%
Let us now introduce the assumption, holding for collisionless of weakly
interacting particles, either neutral or charged, that the Hamiltonian $%
H\left( \mathbf{x}\right) $ can be expressed in the form%
\begin{equation}
H\left( \mathbf{x}\right) =\sum\limits_{i=1,N}H_{i},
\end{equation}%
where $H_{i}=H_{i}\left( \mathbf{x}_{i}\right) .$ This implies that the
particles of the $N$-body system interact mutually only via a mean-field
Hamiltonian force. It follows that the Liouville equation (\ref{Liouville
equation}) admits the factorized solution%
\begin{equation}
\widehat{\rho }^{(N)}(\mathbf{x})=N\prod\limits_{i=1,N}\widehat{p}(\mathbf{x%
}_{i})  \label{factorization}
\end{equation}%
where $\widehat{p}(\mathbf{x}_{i})=\widehat{\rho }^{(1)}(\mathbf{x}_{i})/N$
is the kinetic (1-particle) probability density and $\widehat{\rho }(\mathbf{%
x}_{i})$ is the related kinetic distribution function, which manifestly
obeys the covariant kinetic equation of the form \cite%
{Nicolini2005a,Tessarotto2004}%
\begin{equation}
\frac{ds_{(i)}}{d\tau }\left[ \widehat{\rho }^{(1)}(\mathbf{x}_{(i)}),H_{(i)}%
\right] _{\mathbf{x}_{i}}=0.
\end{equation}
Notice that this equation is independent of $N$, the number of
particles, to be assumed in the sequel as finite.

\section{Observer-related relativistic Gibbs and kinetic entropy}

Let us now introduce the definition for the microscopic entropy $S(\rho
^{(N)})$ appropriate in the context of the present covariant theory. \ The
definition follows by straightforward generalization of the relativistic
kinetic entropy \cite{Nicolini2005a} and the concept of Gibbs entropy in
non-relativistic \cite{Grad 1956,Cercignani 1975} and relativistic \cite%
{Teubner 1961,Israel 1963,De Groot 1980} statistical mechanics.
Thus the concept of Gibbs microscopic entropy can be defined in
analogy to Ref. I, with respect to an observer endowed with proper
time $\tau $
\begin{equation}
S(\rho ^{(N)})=-\int\limits_{\Gamma ^{N}}d\mathbf{x}(s)\prod%
\limits_{j=1,N}\delta (s_{j}-s_{j}(\tau ))\prod\limits_{k=1,N}\delta (\sqrt{%
u_{k\mu }u_{k}^{\mu }}-1)\rho ^{(N)}\ln \rho ^{(N)},
\end{equation}%
Here the notation is as follows: $d\mathbf{x}(s)\equiv \prod\limits_{j=1,N}d%
\mathbf{x}_{j}(s_{j}),$ where for each particle $j=1,N$ the state vectors $%
\mathbf{x}_{i}$ are parameterized with respect to $s_{i}$, with $s_{i}$
denoting the $s$-particle arc length. Finally, $P$ denotes the principal
value of the integral introduced in order to exclude from the integration
domain the subset in which the distribution function vanishes. \ It is
immediate to obtain the relationship between Gibbs and kinetic entropy, $%
S(\rho ^{(1)}),$ \ previously defined in Ref.I and given by the equation

\begin{equation}
S(\rho ^{(N)})=-\int\limits_{\Gamma }d\mathbf{x}_{1}(s)\delta
(s_{1}-s_{1}(\tau ))\delta (\sqrt{u_{1\mu }u_{1}^{\mu }}-1)\rho
^{(1)}\ln \rho ^{(1)},
\end{equation}%
where $\Gamma $ denotes the 1-particle phase space. In fact the condition of
factorization (\ref{factorization}) implies immediately
\begin{equation}
S(\rho ^{(N)})=NS(\rho ^{(1)}).  \label{Gibbs and kinetic entropy}
\end{equation}%
The kinetic entropy can also be written in the equivalent way
\begin{equation}
S(\rho ^{(1)})=-P\int\limits_{\Gamma }d{\mathbf{x}}_{1}(s)\delta
(s_{1}-s_{1}(\tau ))\rho _{1}^{(1)}({\mathbf{x}})\ln \rho
^{(1)}(\mathbf{x}), \label{kinetic entropy}
\end{equation}%
where $\rho _{1}^{(1)}({\mathbf{x}}(s))$ reads%
\begin{equation}
\rho _{1}^{(1)}({\mathbf{x}}(s))=\Theta (r_{1}(s_{1})-r_{\epsilon
}(s_{1}))\delta (\sqrt{u_{1\mu }u_{1}^{\mu }}-1)\rho ^{(1)}({\mathbf{x}}(s)),
\end{equation}%
where $\Theta $ denotes the strong Heaviside function
\begin{equation}
\Theta (a)=\left\{
\begin{array}{ccc}
1 & for & a>0 \\
0 & for & a\leq 0.%
\end{array}%
\right.
\end{equation}%
Equations (\ref{Gibbs and kinetic entropy}) and (\ref{kinetic entropy})
allow to determine immediately the entropy production rate associated to the
Gibbs entropy, which reads
\begin{equation}
\frac{dS(\rho ^{(N)})}{d\tau }=-N\int\limits_{\Gamma ^{-}}d^{3}{\mathbf{r}}%
_{1}d^{3}{\mathbf{p}}_{1}F_{{rr_{\epsilon }}}\delta \left( r_{1}-r_{\epsilon
1}\right) \hat{\rho}^{(1)}\ln \hat{\rho}^{(1)}\equiv \overset{\cdot }{S}_{1}+%
\overset{\cdot }{S}_{2},  \label{entro}
\end{equation}%
where $\Gamma ^{-}$ is the subdomain of phase space in which $n_{\alpha
}u_{i}^{\alpha }\leq 0$ and $F_{{r}{r_{\epsilon }}}$ is the characteristic
integrating factor
\begin{equation}
F_{{r}_{i}{r_{\epsilon i}}}\equiv \frac{ds_{i}(\tau )}{d\tau }\left( \frac{%
dr_{i}}{ds_{i}}-\frac{dr_{\epsilon i}}{ds_{i}}\right) .
\end{equation}%
It follows that $\frac{dS(\rho ^{(N)})}{d\tau }$and can be interpreted as
the \textit{average of the entropy flux across the capture surface} $\gamma
. $ Moreover, $\overset{\cdot }{S}_{1}$ and $\overset{\cdot }{S}_{1}$ denote
respectively the contributions to entropy production rate%
\begin{equation}
\overset{\cdot }{S}_{1}=-NP\int\limits_{\Gamma ^{-}}d^{3}{\mathbf{r}}%
_{1}d^{3}{\mathbf{p}}_{1}\frac{ds_{1}(\tau )}{d\tau }\frac{dr_{1}}{ds_{1}}%
\delta \left( r_{1}-r_{\epsilon 1}\right) \hat{\rho}^{(1)}\ln \hat{\rho}%
^{(1)},
\end{equation}%
\begin{equation}
\overset{\cdot }{S}_{2}=NP\int\limits_{\Gamma ^{-}}d^{3}{\mathbf{r}}%
_{1}d^{3}{\mathbf{p}}_{1}\frac{ds_{1}(\tau )}{d\tau }\frac{dr_{\epsilon 1}}{%
ds_{1}}\delta \left( r_{1}-r_{\epsilon 1}\right) \hat{\rho}^{(1)}\ln \hat{%
\rho}^{(1)}.
\end{equation}

We stress that here by construction $\frac{ds_{1}(\tau )}{d\tau }>0$ [see
Eq.(\ref{assumption})] and $\frac{dr_{1}}{ds_{1}}<0,$ $\frac{dr_{1}}{ds_{1}}$
denoting the "radial" velocity of infalling matter on the surface $\gamma ,$
while there results $\frac{dr_{\epsilon 1}}{ds_{1}}<0,=0$ or $>0,$ being $%
\frac{dr_{\epsilon 1}}{ds_{1}}$the local velocity of the surface $\gamma ,$
respectively for contracting, stationary and expanding event horizons.
However, the signs of $\overset{\cdot }{S}_{1}$ and $\overset{%
\cdot }{S}_{2}$ are generally not defined, unless further
assumptions are taken into account. Notice that, in analogy to the
Bekenstein position (\ref{Beck}), $\overset{\cdot }{S}_{1}$and $\overset{%
\cdot }{S}_{2}$ denote the contributions to the entropy flux
carried by incoming matter and by the BH due to the motion of the
event horizon,
therefore they can be at least qualitatively related in the following way:%
\begin{eqnarray}
\overset{\cdot }{S}_{1} &\rightarrow &\overset{\cdot }{S}, \\
\overset{\cdot }{S}_{2} &\rightarrow &\overset{\cdot }{S}_{bh},  \notag
\end{eqnarray}%
being $S$ and $S_{bh}$ the contributions to Bekenstein entropy (\ref{Beck}).
In particular, it interesting to remark that both $\overset{\cdot }{S}_{1}$%
and $\overset{\cdot }{S}_{2}$ \emph{result by construction proportional to} $%
A,$ the area of the event horizon, a conclusion which appears in
qualitative agreement with estimate for the BH Boltzmann entropy
given above for $S_{bh}$.

\section{H-theorem for the Gibbs entropy}

Let us now introduce the assumption that the total number of
particles is finite, but suitable large ($N\gg 1$). In such a
case, it is possible to determine the signs of $\overset{\cdot
}{S}_{1}$, which contributes to the entropy production rate
(\ref{entro}). Indeed it is possible to
prove that there results $\overset{\cdot }{S}_{1}> 0$ while $\overset{%
\cdot }{S}_{2}$ has not a definite sign. In addition, thanks to
the results given in the previous section, in particular the
relationship between the Gibbs and kinetic entropies, $S(\rho
^{(N)})$ and $S(\rho ^{(1)}),$ specified by Eq.(\ref{Gibbs and
kinetic entropy}),\ the following H-theorem holds for \ $S(\rho
^{(N)}),$

\begin{equation}
\frac{dS(\rho ^{(N)})}{d\tau }\equiv \overset{\cdot }{S}_{1}+\overset{\cdot }%
{S}_{2}\geq 0.
\end{equation}

 Moreover, we notice that the support of
the kinetic distribution function, i.e., the subset of $\Gamma
^{-}$ in which the kinetic distribution function is non negative,
results always compact. This condition is as a direct consequence
of the Assumption 2 here considered, implying
that the energy of the falling particles reaching the surface $%
\gamma $ cannot become infinite.

Therefore, denoting $\Omega $ the subset of \ $\Gamma ^{-}$ in which the
kinetic distribution function $\hat{\rho}^{(1)}$ is non-zero, in the
complementary set $\Gamma ^{-}\setminus \Omega ,$ the kinetic distribution
function $\rho ^{(1)}=Np^{(1)}$ (being $p^{(1)}$ the kinetic probability
density) results identically zero. Thanks to Assumption 2, it follows that
such a set results necessarily bounded. Therefore, so that the following
majorization holds
\begin{equation}
\frac{dS(\rho ^{(1)})}{d\tau }\geq P{\int\limits_{\Omega }}d^{3}{\mathbf{r}}%
d^{3}{\mathbf{p}}\left\vert F_{{r}{r_{\epsilon }}}\right\vert \delta \left(
r-r_{\epsilon }\right) \left[ Np^{(1)}-1\right] .  \label{vel}
\end{equation}%
Thus, letting
\begin{equation}
M_{\delta }\equiv {\int\limits_{\Omega }}d^{3}{\mathbf{r}}d^{3}{\mathbf{p}}%
\left\vert F_{{r}{r_{\epsilon }}}\right\vert \delta \left( r-r_{\epsilon
}\right)
\end{equation}%
and imposing that $N\gg 1$ be sufficiently large to satisfy the inequality
\begin{equation}
\frac{dS(\rho ^{(1)})}{d\tau }\geq \dot{S}\equiv N\inf \left\{ \int_{\Omega
}d^{3}\mathbf{r}n_{0}V_{r}^{eff}\right\} -M_{\delta }\geq 0  \label{essep}
\end{equation}%
the thesis of the H-theorem is reached provided we assume $\inf \left\{ \int
d^{3}\mathbf{r}n_{0}V_{r}^{eff}\right\} >0,$ a condition consistent with the
requirement of a non isolated BH surrounded by matter. \ In the previous
equation we have introduced the additional notation
\begin{equation}
P\int\limits_{\Gamma ^{-}}d^{3}{\mathbf{r}}_{1}d^{3}{\mathbf{p}}_{1}\frac{%
ds_{1}(\tau )}{d\tau }\frac{dr_{1}}{ds_{1}}\delta \left( r_{1}-r_{\epsilon
1}\right) \hat{\rho}^{(1)}=N\int_{\Omega }d^{3}\mathbf{r}n_{0}V_{r}^{eff},
\end{equation}%
being $n_{0}$ the number density. We stress that this result generalized the
H-theorem given in Ref.I, since it applies also to Gibbs entropy. The result
holds for classical BH having, in principle, arbitrary shape of the event
horizon and even in the presence of a contracting event horizon (which might
by produced, for example, by star implosions). \ The present theory appear
therefore potentially relevant for a realistic detailed analysis of the BH
thermodynamical properties.

\section{Conclusions}

In this paper a macroscopic statistical description has been
adopted for classical matter around black holes.  Matter in the
immediate vicinities of a BH event horizon has been modelled by a
weakly interacting relativistic gas $S_{N}.$ Its dynamics results
described by the relativistic Lioville equation, while the
presence of the BH event horizon is taken into account by treating
it as a classical absorbing porous wall.

By assuming that Hamiltonian dynamics takes into account only mean
field interactions between particles, the connection with the
kinetic treatment of Ref.I can be immediately established. As a
consequence, an H-theorem valid for the Liouville equation can be
established on rigorous grounds which applies to every space time
geometry and to the case of contracting horizon.

\section{Acknowledgments}

Work developed with the support of Ministero dell'Istruzione, dell'Universit%
\`{a} e della Ricerca (MIUR), via the Programma PRIN 2004:
"Modelli della teoria cinetica matematica nello studio dei sistemi
complessi nelle scienze applicate".

\section{Appendix: relativistic N-body Liouville theorem}

Let us assume that the point particles constitute an isolated N-body system (%
\emph{assumption} $\alpha $) obeying the relativistic Hamiltonian equations
of motion%
\begin{align}
\frac{d\mathbf{x}_{i}(s_{i})}{ds_{i}}& =\mathbf{X}_{i},
\label{canonical equations} \\
\mathbf{x}_{i}\mathbf{(s}_{io}\mathbf{)}& =\mathbf{x}_{oi},
\label{initial conditions}
\end{align}%
which implies that all the vector fields $\mathbf{X}_{i}$ ($i=1,N$) are
conservative, i.e., for $i=1,N:$%
\begin{equation}
\frac{\partial }{\partial \mathbf{x}_{(i)}}\cdot \mathbf{X}_{i}=0.
\label{conservative system}
\end{equation}%
Introducing the parametrization in terms of the observer's time $\tau $ and
requiring that the functions $s_{i}=s_{i}(\tau )$ are strictly monotonic,
the equations of motion can be written in the symbolic form it follows%
\begin{equation}
\frac{d\mathbf{x}_{i}(s_{i})}{d\tau }=\frac{ds_{(i)}}{d\tau }\frac{d\mathbf{x%
}_{i}(s_{i})}{ds_{i}}=\frac{ds_{(i)}}{d\tau }\mathbf{X}_{i}.
\end{equation}%
Here $\mathbf{x}=\left\{ \mathbf{x}_{1},....,\mathbf{x}_{N}\right\}
_{i}\equiv \left\{ y_{1},.....,y_{8N}\right\} ,$ $\mathbf{X}=\left\{ \mathbf{%
X}_{1},....,\mathbf{X}_{N}\right\} _{i}\equiv \left\{
Y_{1},.....,Y_{8N}\right\} ,$ where by assumption $\alpha $ the vector field
$\mathbf{X}$ depends only on the (local or retarded) states of the particles
forming the $N$-body system. As a consequence, let us denote
\begin{equation}
\mathbf{x}(\tau )\equiv \mathbf{x}(s(\tau ))\equiv \left\{ \mathbf{x}%
_{1}(s_{1}(\tau )),...,\mathbf{x}_{N}(s_{N}(\tau ))\right\}
\end{equation}%
the solution of the initial value problem (\ref{canonical equations})-(\ref%
{initial conditions}) and $\mathbf{x(}\tau _{o}\mathbf{)}\equiv \mathbf{x}%
(s(\tau _{o}))\equiv \left\{ \mathbf{x}_{1}(s_{1}(\tau _{o})),...,\mathbf{x}%
_{N}(s_{N}(\tau _{o}))\right\} =\mathbf{x}_{0}$ the initial condition$,$
where $\mathbf{x}(\tau )$ and $\mathbf{x(}\tau _{o}\mathbf{)}$ denote the $%
N- $body system states as seen by the observer $O,$ respectively at times $%
\tau ,\tau _{o}.$ \ The previous assumptions for the phase-mapping
\begin{equation}
\mathbf{x}_{o}\rightarrow \mathbf{x}(\tau )=\mathbf{\chi (x}_{o},\tau
_{o},\tau )
\end{equation}%
imply the following theorem:

\subsubsection{THM. - Relativistic N-body Liouville theorem}

\emph{For arbitrary }$x_{o}\in \Gamma $\emph{\ and }$\tau _{o},\tau \in
I\subseteq
\mathbb{R}
$\emph{\ there results }%
\begin{equation}
\left\vert \frac{\partial \mathbf{x}(\tau )}{\partial \mathbf{x}_{o}}%
\right\vert =1.
\end{equation}

In fact, for $N>1$ the time derivative of the Jacobian $\left\vert \frac{%
\partial \mathbf{x}(\tau )}{\partial \mathbf{x}_{o}}\right\vert $ reads

\begin{equation}
\frac{d}{d\tau }\left\vert \frac{\partial \mathbf{x}(\tau )}{\partial
\mathbf{x}_{0}}\right\vert =\sum\limits_{i=1,8N}\frac{ds_{i}}{d\tau }%
\left\vert \frac{\partial \left( y_{1}(\tau )\mathbf{,....,}\frac{dy_{i}}{%
ds_{i}},...,y_{8N}(\tau )\right) }{\partial \mathbf{x}_{0}}\right\vert .
\end{equation}%
Hence by the chain rule%
\begin{equation*}
\frac{d}{d\tau }\left\vert \frac{\partial \mathbf{x}(\tau )}{\partial
\mathbf{x}_{0}}\right\vert =
\end{equation*}%
\begin{equation}
=\sum\limits_{i=1,8N}\frac{ds_{i}}{d\tau }\sum\limits_{r=1,8N}\frac{%
\partial Y_{i}}{\partial y_{r}(\tau )}\left\vert \frac{\partial \left(
y_{1}(\tau )\mathbf{,....,}\frac{dy_{i}}{ds_{i}},...,y_{8N}(\tau )\right) }{%
\partial \mathbf{x}_{0}}\right\vert =
\end{equation}%
\begin{equation*}
=\left\vert \frac{\partial \mathbf{x}(\tau )}{\partial \mathbf{x}_{0}}%
\right\vert \sum\limits_{i=1,8N}\frac{ds_{i}}{d\tau }\frac{\partial Y_{i}}{%
\partial y_{i}}\equiv \left\vert \frac{\partial \mathbf{x}(\tau )}{\partial
\mathbf{x}_{0}}\right\vert \sum\limits_{i=1,N}\frac{ds_{i}}{d\tau }\frac{%
\partial }{\partial \mathbf{x}_{i}}\cdot \mathbf{X}_{i},
\end{equation*}%
and thanks to the condition of conservation (\ref{conservative system})

\begin{equation}
\frac{d}{d\tau }\left\vert \frac{\partial \mathbf{x}(\tau )}{\partial
\mathbf{x}_{0}}\right\vert =0,
\end{equation}%
which implies the thesis. \textbf{c.v.d.}

\end{document}